\documentclass{article}

\usepackage{arxiv}
\usepackage[utf8]{inputenc}
\usepackage[T1]{fontenc}
\usepackage{amsmath,amssymb}
\usepackage{graphicx}
\usepackage{booktabs}
\usepackage{longtable}
\usepackage{array}
\usepackage{calc}
\usepackage{hyperref}
\usepackage{url}
\usepackage{xcolor}
\usepackage{tabularx}
\usepackage{float}
\usepackage{listings}
\usepackage{tikz}
\usetikzlibrary{arrows.meta, positioning, decorations.pathreplacing}
\usepackage{needspace}

\providecommand{\tightlist}{%
  \setlength{\itemsep}{0pt}\setlength{\parskip}{0pt}}

\makeatletter
\@namedef{ftype@none}{8}
\makeatother

\lstdefinestyle{pseudocode}{
  basicstyle=\small\ttfamily,
  backgroundcolor=\color{gray!8},
  frame=single,
  rulecolor=\color{gray!40},
  framesep=6pt,
  xleftmargin=8pt,
  xrightmargin=8pt,
  aboveskip=10pt,
  belowskip=10pt,
  mathescape=true,
  columns=fullflexible,
  keepspaces=true,
  literate={->}{$\to$}2 {=>}{$\Rightarrow$}2 {++}{$\mathbin{+\!\!+}$}2 {bot}{$\bot$}3,
}

\title{Send: Objects, History, and Transactions in a Single-Verb Kernel}
\author{
  Christopher Goes \\
  \texttt{cwgoes@pluranimity.org}
}
\date{May 2026}

\begin{document}

\maketitle

\begin{abstract}
Multi-party object coordination --- across object-capability systems, smart-contract platforms, distributed actors, and event-sourced architectures --- is shaped by six structural properties: authenticated provenance, opaque encapsulation, atomic multi-object commit, deterministic replay, immutable history, and history-derived state. Existing systems compose subsets via separate layered mechanisms (RPC, capability ACLs, transaction coordinators, event journals, vat boundaries); each layer is well-studied but the combination is fragile. We present a minimal kernel which makes them jointly compatible.

Our kernel is built from s-expressions, a uniform \texttt{send} interface, transactions, and one primitive object distinction: \emph{ephemeral} (caller's context inherited) vs.~\emph{persistent} (context switches to the target's kernel-assigned identity and append-only log). The kernel structurally classifies every send target into one of six cases without input from the caller --- uniform caller interface, intensional kernel dispatch.

Under kernel-faithful trust (the kernel runs its semantics as specified), this design holds all six properties as \emph{kernel-level} against arbitrary programs --- the kernel's transition function refuses states violating them --- in the precise sense defined in §4.4 (the §5 instantiation realises identity-pattern opacity at user-level under its sequential allocator; a hash-allocator variant restores it to kernel-level --- §5.6). Opacity \emph{against the operator} additionally requires operator-faithful trust (the operator accesses logs only via \texttt{recall} and does not censor or reorder transactions); under kernel-faithful alone, five of six guarantees survive an unconstrained operator. Append-only logs underpin immutability, replay, and history-derived state; kernel-controlled persistent dispatch yields authenticated provenance and opacity; transactions deliver atomic coordination. Opacity has one structural exception: co-participants in a transaction observe its commit-or-abort outcome (Proposition 4), an intentional consequence of atomic coordination rather than a leak. Operator-adversarial deployments can be realized with a cryptographic compiler (§7 Case Study 2, Viaduct).
\end{abstract}

\section{Introduction}

Multi-party object coordination is shaped by six structural concerns --- authenticated provenance\cite{ref9}, opaque encapsulation\cite{ref15}, atomic multi-object commit\cite{ref17}, deterministic replay\cite{ref23}, immutable history\cite{ref19}, and history-derived state\cite{ref18} --- each individually motivated by patterns in this domain. We use \emph{multi-party object coordination} throughout to refer to the broad domain spanning object-capability systems, smart-contract platforms, distributed actor frameworks, and event-sourced architectures, wherever multiple principals interact through encapsulated objects whose state must be jointly authenticated, atomically updated, and replayable from a shared record. We propose this as a category-name for an implicitly-shared design space: no prior literature treats these system classes as one, but the six requirements above span all of them, which makes the cross-class comparison §8 draws coherent. These concerns are conventionally provided for by separate, layered mechanisms (RPC, capability ACLs, transaction coordinators, event journals, and vat boundaries), with state held in mutable snapshots that must be kept in sync with the event log to remain auditable. Each layer is well-studied in isolation, but the combination is fragile: interfaces multiply, cross-layer invariants are easy to violate, and security properties leak across layer boundaries. What design provides all six simultaneously, not as separate layered mechanisms but as kernel-level properties?

Our kernel is designed as a \emph{target} representation, not a surface language: a structural specification that source object languages compile to, and that cryptographic compilers (Viaduct-style\cite{ref24}, §7 Case Study 2) take as a specification of what must remain opaque, atomic, and replay-verifiable under adversarial operators. The §5 instruction set is correspondingly chosen for computational sufficiency, not human readability; usable surface languages are downstream work.

The kernel rests on three commitments developed in §2:

\begin{itemize}
\tightlist
\item
  \emph{Ephemeral/persistent dispatch axis.} A send target's structural classification determines whether execution context switches (persistent) or is inherited (ephemeral). The cleanest demonstration is §7 Case Study 1, where the only difference between two otherwise-identical traces is this classification.
\item
  \emph{Send as uniform cross-object interface.} The sole cross-object verb, classified by the kernel into six structural cases; intra-object computation uses additional kernel-local instructions (§5) that do not cross object boundaries.
\item
  \emph{State as history.} A persistent object's state is what its program derives from its append-only log.
\end{itemize}

These commitments, combined with the transaction mechanism, yield a kernel providing these six properties. The kernel's distinctive ergonomic move is \emph{interface} unification: the caller uses one verb, and the kernel performs an intensional six-way structural classification of the target --- encapsulating distinctions which prior systems expose as separate caller-selected mechanisms (Self's parent slots, Smalltalk's super-sends, EVM's \texttt{DELEGATECALL}).

\textbf{Kernel-level enforcement.} A property is \emph{kernel-level} for a system if the transition function guarantees that it holds, regardless of program input; \emph{user-level} if well-formed programs can violate it. Smalltalk's \texttt{instVarAt:} makes opacity user-level; Akka Persistence's non-deterministic event handlers make deterministic replay user-level. We show that this kernel achieves uniform kernel-level enforcement across all six properties (Definition + Theorem 3 + Proposition 4, §4.4) under these definitions; operator-faithful trust is required for opacity (see ``Scope and trust'' below). Each of the six properties is well-studied in isolation, and many existing systems realise subsets via composed mechanisms, but alternative designs distribute the kernel/user-level boundary differently. The row-by-row contrast in §8 makes the comparison precise.

\textbf{Scope and trust.} The kernel is positioned with respect to three trust regimes, distinguished by what is assumed about the deployment operator.

\emph{Kernel-faithful}: the kernel implementation runs its semantics as specified. All six structural properties hold against arbitrary programs. Opacity \emph{against the operator} additionally requires the operator-faithful conditions below.

\emph{Operator-faithful}: in addition to kernel-faithfulness, the deployment operator satisfies three sub-faithfulness conditions: (a) \emph{commit-faithfulness}: every well-formed transaction submitted to the kernel is committed (no selective censorship); (b) \emph{order-faithfulness}: transactions are ordered consistently with the wrapping protocol's \(T\) (the operator does not reorder \(T\) beyond the conflict-serialisable equivalences §6.3 permits); (c) \emph{read-faithfulness}: the operator stores \(K\) but does not leak its contents into in-kernel program execution outside the kernel's defined read interfaces (own-log via position 3; cross-object only via send). Under operator-faithfulness, opacity holds against all programs; the operator is the only entity that can read a persistent object's full \(K\). Opacity is \emph{at the read interface}: atomic transactional coupling necessarily exposes co-participation across receivers (Proposition 4, §4.4).

\emph{Operator-adversarial}: the operator may violate any of (a)--(c). The kernel does not solve this regime; cryptographic-compiler composition (§7 Case Study 2, Viaduct\cite{ref24}) is the standard remedy, and the kernel serves as the structural target the compiler instantiates against adversarial operators. Case Study 2 sketches label-derivation from object identities for the static-allocation case; dynamic allocation is future work.

The kernel itself is not a distributed system; the wrapping protocol's properties (consistency, availability, partition-tolerance trade-offs\cite{ref29}; liveness under asynchrony) are scope-conditional on the wrapper choice (§6.2), not kernel concerns. The kernel exposes strict serializability under sequential processing with a real-time-faithful wrapper (§3); §6.3 discusses conflict-serializable scheduling and its parallelism floor. The kernel assumes program termination (§3.4); resource-bounding is composition-layer scope, and under the compilation-target framing becomes a compiler obligation. Recovery is discussed in §6.4.

\emph{External principals.} \texttt{execute} assigns caller \(= a_{\mathrm{ext}}\) for top-level transactions; the wrapping protocol (signed transactions, authenticated channels, consensus-attached identity) authenticates which external party submitted each transaction. Multi-party scenarios represent external principals in-kernel as persistent proxy objects (the §7 case studies do this for bidders, buyers, and sellers), with the wrapper routing each authenticated submission to its designated proxy; the kernel's authenticated-provenance claim is over these intra-kernel identities, conditional on this wrapper discipline.

The kernel's essential semantics are parameterised. Section 2 presents the invariant core informally. Section 3 formalises it. Section 4 derives the structural properties. Section 5 gives one concrete instantiation. Section 6 addresses locality, composition modes (replication, sharding), conflict-serializable scheduling, and log retention. Section 7 shows that common patterns are encodable within the kernel. Section 8 positions the kernel against prior systems. Section 9 outlines open directions.

\section{Essential Semantics}

Invariant core of the kernel, informally; §3 formalises.

\textbf{Data.} One data type: s-expressions. Every s-expression is an object, meaning it can be the target of a send. No structure beyond being an s-expression is required to be an object.

\textbf{Send.} The cross-object interface. A program specifies a target and a message; the kernel determines what happens based on the target's classification. The sender's interface is uniform: it does not select which kind of send to perform. The kernel classifies every target into one of six structurally distinct cases plus an invalid catch-all. Two are primary modes: persistent object (context switch to the target's program) and ephemeral object (runs with inherited context). Four are boundary cases: built-in (pure computation), kernel (object creation), external target (fire-and-forget), and special pair form (instantiation-defined dispatch on certain pair shapes; §5.3). A target matching none of the six aborts.

\textbf{Ephemeral vs.~persistent objects.} The one boundary. Ephemeral objects are transparent: no identity, no history, no boundary. Persistent objects are created by sending a program to the kernel. Each has an opaque identity (an atom), an append-only log of authenticated interactions, and a program. No one can read another object's log or program --- the kernel enforces this absolutely. Inside is invisible from outside; every crossing from outside to inside is authenticated, logged, and bilateral.

\emph{Design-space observation: the ephemeral/persistent endpoints.} The ephemeral/persistent axis bundles three sub-distinctions --- identity (named vs.~anonymous), logging (logged vs.~unlogged), context (switching vs.~inheriting). Under this paper's design premises --- the six structural properties of §1, enforced as kernel-level guarantees with the specific definitions §4.4 makes precise --- only two of the eight (id, log, ctx) configurations satisfy all six simultaneously: \emph{persistent} (named, logged, switching) and \emph{ephemeral} (anonymous, unlogged, inheriting). The kernel takes these as the dispatch axis. The remaining six configurations each fail at least one property \emph{under these definitions}:

\noindent
\small
\begin{tabularx}{\columnwidth}{@{} l X @{}}
\toprule
Configuration & Property unsatisfied (§4.4 def.) \\
\midrule
named, logged, inheriting & Provenance: log on receiver but caller's identity runs \\
named, unlogged, switching & Replay, immutability, history-derived state (no log) \\
named, unlogged, inheriting & Replay, immutability, history-derived state \\
anonymous, logged, switching & Provenance: log without identity to attribute to \\
anonymous, logged, inheriting & Provenance: log without owner; delegation accountability broken \\
anonymous, unlogged, switching & Cannot switch to a nameless target \\
\bottomrule
\end{tabularx}
\normalsize

Other configurations are valid for systems with different commitments --- e.g., named-unlogged-switching is traditional mutable OO (Smalltalk, Self), forfeiting replay and history-derived state for in-place mutation. Our claim is not that other configurations are impossible but that the two endpoints above are what permit all six properties to hold simultaneously as kernel-level guarantees.

\textbf{State as history.} Persistent state is append-only: a persistent object's state is whatever its program derives from its interaction log. Each log entry records \([\mathrm{caller}, \mathrm{msg}]\) --- the authenticated identity of the sender and the message. The first entry is a birth record placed by the kernel at creation time.

\textbf{Transactions.} The sole coordination mechanism. Everything within one submission to the kernel is a single atomic transaction --- all effects commit or none do. Positive coupling: bundled sends co-occur. Negative coupling: any receiver can abort, destroying the entire bundle. Details in Section 4.

\textbf{Parameterisation.} The essential semantics are parameterised over an atom set, a set of built-in objects, a program model, and a log encoding. Section 3 defines the constraints on each parameter and the formal dispatch rules.

\section{Formal Model of Essential Semantics}

The mathematical core, parameterised over \(A\), \(B\), \(\Pi\), interpreter, and encoding. Three definitions: \(\mathrm{send}\) (six-case dispatch), \(\mathrm{run}\) (program interpreter), \(\mathrm{execute}\) (transaction commit). Readable as small-step operational semantics or as a transition function on the global state \(K\).

\subsection{Data Domain}

\[\mathcal{S} = A + \mathcal{S} \times \mathcal{S}\]

Structural equality:

\[n = m \iff n \text{ and } m \text{ are the same element of } A\] \[[a,b] = [c,d] \iff a = c \text{ and } b = d\] \[n \neq [a,b] \quad \text{for all } n \in A,\; [a,b] \in \mathcal{S} \times \mathcal{S}\]

\subsection{State}

\[K \in (A \times A \times \mathcal{S})^*\]

The global persistent state --- all persistent object histories in a single append-only sequence. Each entry \((n, c, m)\) records that persistent object \(n\) received message \(m\) from caller \(c\). The caller \(c\) is always an atom --- either a persistent object's identity or \(a_{\mathrm{ext}}\) --- because the kernel-controlled caller (position 4) is always an element of \(A\).

Derived operations (projections of \(K\) onto individual persistent objects):

\[\mathrm{log}(n, K) = [(c, m) \mid (n, c, m) \in K]\] \[\mathrm{exists}(n, K) \iff n \in A \setminus \{a_{\mathrm{kernel}}\} \text{ and } \exists\, c.\ (a_{\mathrm{kernel}}, c, n) \in K\] \[\mathrm{Pers}(K) = \{n \in A : \mathrm{exists}(n, K)\}\] \[\mathrm{program}(n, K) = \mathrm{second}(\mathrm{log}(n, K)[0])\]

Parameters (supplied by the instantiation):

\[\mathrm{encode} : (A \times \mathcal{S})^* \to \mathcal{S} \qquad \Pi : \mathcal{S} \times \mathcal{S} \to \mathcal{S} + \{\bot\}\] \[\mathrm{Ext}, \mathrm{Eph} \subseteq \mathcal{S} \times \mathcal{S} \qquad B \subseteq A\] \[\mathrm{builtin} : A \times \mathcal{S} \to \mathcal{S} + \{\bot\} \qquad \mathrm{alloc} : (A \times A \times \mathcal{S})^* \to A\]

\(\Pi\) is the \emph{instantiation-defined pair-form} dispatch: non-\(\bot\) for pairs matching an instantiation-specified special form (e.g., §5.3 pair closures). The instantiation must ensure: (i) \(\Pi\)-matched pairs, \(\mathrm{Ext}\), and \(\mathrm{Eph}\) partition \(\mathcal{S} \times \mathcal{S}\); (ii) \(\mathrm{builtin}(n, m) = \bot\) for all \(n \notin B\) (built-ins are defined only on \(B\); behaviour on ill-formed arguments may abort); (iii) \(\mathrm{alloc}\) returns a fresh atom outside \(B \cup \{a_{\mathrm{kernel}}, a_{\mathrm{ext}}\} \cup \mathrm{Pers}\); (iv) \(\mathrm{encode}\) is injective; (v) the program model is pure (no inspection of \(K\), \(T\), time, or randomness).

Transaction effects (accumulated during execution):

\[\Delta \in (A \times A \times \mathcal{S})^*, \qquad \Xi \in (A \times \mathcal{S} \times \mathcal{S})^*\]

\(\Delta\) holds pending entries on persistent objects' logs; \(\Xi\) holds pending external sends.

Read-your-writes: all lookups during a transaction use \(K \mathbin{+\!\!+} \Delta\) --- each persistent object sees its committed history plus pending entries from completed prior invocations.

Transaction log (system-level boundary record):

\[T \in (\mathcal{S} \times (\mathcal{S} + \{\bot\}) \times (A \times \mathcal{S} \times \mathcal{S})^*)^*\]

Each entry \((\mathrm{tx}, v, \Xi)\) records a submitted transaction \(\mathrm{tx}\), its result \(v\) (or \(\bot\) on abort), and the external sends \(\Xi\) dispatched on commit (empty on abort).

\subsection{Send}

\[\mathrm{send} : \mathcal{S} \times \mathcal{S} \times A \times \mathcal{S} \times A \times K \times \Delta \times \Xi \to (\mathcal{S} \times \Delta \times \Xi) + \{\bot\}\]

All cases, parameterised. Cases in order; first match wins.

\begin{align*}
\mathrm{send}&(t, m, s, l, c, K, \Delta, \Xi) = \\
&\begin{cases}
\bot & t \in B,\; \mathrm{builtin}(t, m) = \bot \\
(\mathrm{builtin}(t, m),\; \Delta,\; \Xi) & t \in B,\; \mathrm{builtin}(t, m) \ne \bot \\
\mathrm{send}_{\mathrm{kern}}(s, m, K, \Delta, \Xi) & t = a_{\mathrm{kernel}} \\
\mathrm{send}_{\mathrm{pers}}(t, m, s, K, \Delta, \Xi) & t \in \mathrm{Pers}(K \mathbin{+\!\!+} \Delta) \\
(\Pi(t, m),\; \Delta,\; \Xi) & t \in \mathcal{S} \times \mathcal{S},\; \Pi(t, m) \ne \bot \\
(a_{\mathrm{ext}},\; \Delta,\; \Xi \cdot (s, t, m)) & t \in \mathrm{Ext} \\
\mathrm{run}([t, m, s, l, c],\; t,\; K,\; \Delta,\; \Xi) & t \in \mathrm{Eph} \\
\bot & \text{otherwise}
\end{cases}
\end{align*}

with the two non-trivial cases factored as named helpers:

{\small
\begin{align*}
\mathrm{send}_{\mathrm{pers}}(t, m, s, K, \Delta, \Xi) &= \begin{cases}
\bot & \text{if } r = \bot \\
(v,\; \Delta' \cdot (t, s, m),\; \Xi') & \text{if } r = (v, \Delta', \Xi')
\end{cases} \\
&\text{where } p = \mathrm{program}(t, K \mathbin{+\!\!+} \Delta), \\
&\quad h = \mathrm{encode}(\mathrm{log}(t, K \mathbin{+\!\!+} \Delta)), \\
&\quad r = \mathrm{run}([p, m, t, h, s], p, K, \Delta, \Xi) \\
\mathrm{send}_{\mathrm{kern}}(s, m, K, \Delta, \Xi) &= (\mathit{id},\; \Delta',\; \Xi) \\
&\text{where } \mathit{id} = \mathrm{alloc}(K \mathbin{+\!\!+} \Delta), \\
&\quad \Delta' = \Delta \cdot (a_{\mathrm{kernel}}, s, \mathit{id}) \cdot (\mathit{id}, s, m)
\end{align*}
}

Cases in priority order; first match wins.

\begin{itemize}
\tightlist
\item
  \textbf{Built-in} (\(t \in B\)): Pure computation. No effect on \(\Delta\) or \(\Xi\).
\item
  \textbf{Kernel} (\(t = a_{\mathrm{kernel}}\)): Object creation. Allocates fresh identity. Two pending entries: creation record on kernel's log, birth record \([\mathrm{creator}, \mathrm{program}]\) on new object's log.
\item
  \textbf{Persistent} (\(t \in \mathrm{Pers}(K \mathbin{+\!\!+} \Delta)\)): Context switch. Target's program runs with self \(= t\), log \(= h\), caller \(= s\). The pending entry \((t, s, m)\) is appended to \(\Delta'\) after the target's program returns; \(h\) thus encodes history up to but not including the current message. Computing \(\mathrm{Pers}\) over \(K \mathbin{+\!\!+} \Delta\) admits dispatch to objects created earlier in the same transaction.
\item
  \textbf{Special pair form} (\(\Pi(t, m) \ne \bot\)): Pure pair-form dispatch defined by the instantiation (e.g., §5.3 pair-closure rule). No effect on \(\Delta\) or \(\Xi\).
\item
  \textbf{External} (\(t \in \mathrm{Ext}\)): Fire-and-forget. Returns \(a_{\mathrm{ext}}\). Appends to \(\Xi\).
\item
  \textbf{Ephemeral} (\(t \in \mathrm{Eph}\)): Inherited context. No persistent effects. Applies to pairs not matched by \(\Pi\) or \(\mathrm{Ext}\).
\item
  \textbf{Otherwise}: Abort.
\end{itemize}

\emph{Disjointness.} \(B\), \(\{a_{\mathrm{kernel}}\}\), and \(\mathrm{Pers}(K \mathbin{+\!\!+} \Delta)\) partition the relevant atoms by the instantiation's allocation discipline; pairs are partitioned per §3.2; the catchall covers any remaining target.

\emph{Log ordering.} Entries are appended in \emph{completion order}: \((t, s, m)\) is added to \(\Delta'\) after the target returns. Re-entrant sends to the same target appear before the outer entry. This is forced by the position-3 invariant: the receiver's log visible to its program at \(m\)'s execution must not include \(m\)'s own entry. Trigger-order is exposed externally via \(T\) (which records each \(\mathrm{tx}_i\) as a single unit). Throughout this paper, ``history-derived state'' refers to the completion-ordered log-replay sense; trigger-order recovery, when needed, uses \(T\) or explicit message-level sequence threading.

\subsection{Interpreter}

\[\mathrm{run} : \mathcal{S}^* \times \mathcal{S} \times K \times \Delta \times \Xi \to (\mathcal{S} \times \Delta \times \Xi) + \{\bot\}\]

The program model must support:

\begin{itemize}
\tightlist
\item
  \textbf{Push values} onto the execution context.
\item
  \textbf{Send} --- invoke \(\mathrm{send}\) on the top two values with the current context.
\item
  \textbf{Recall} --- access context positions: program (0), message (1), self (2), log (3), caller (4).
\item
  \textbf{Quote} --- push a value without interpretation.
\item
  \textbf{Abort} --- produce \(\bot\), which propagates through all enclosing computations.
\end{itemize}

The execution context's three kernel-controlled values --- self (position 2), log (position 3), caller (position 4) --- switch on persistent send, are inherited on ephemeral. Concrete instruction set is an instantiation choice. The model assumes program termination; resource-bounding is composition-layer scope (§1 ``Scope and trust'').

\subsection{Transaction}

\[
\mathrm{execute}(K, [p, i]) = \begin{cases}
(\bot,\; K,\; \varepsilon) & \text{if } r = \bot \\
(v,\; K \mathbin{+\!\!+} D,\; X) & \text{if } r = (v, D, X)
\end{cases}
\]

where \(r = \mathrm{run}([p, i, a_{\mathrm{ext}}, \mathrm{encode}(\varepsilon), a_{\mathrm{ext}}], p, K, \varepsilon, \varepsilon)\).

The program runs unprivileged: self \(= a_{\mathrm{ext}}\), log \(= \mathrm{encode}(\varepsilon)\), caller \(= a_{\mathrm{ext}}\). It can create objects, send to existing objects, and compute --- but cannot read the kernel's log or enumerate objects. Success: pending entries commit, externals dispatched. Abort: \(K\) unchanged, no externals.

\textbf{Terminology.} \emph{Transaction} refers to a single submission \([p, i]\) running \(\mathrm{execute}\) to completion or abort: a one-shot stored-procedure invocation in the Calvin\cite{ref25} sense, not a multi-round-trip database transaction. \emph{Serializable} refers to the standard ANSI/Adya isolation level\cite{ref28}; sequential processing (§3) realises \emph{strict serializability} under a wrapper that totally orders submissions consistently with their wallclock-completion times (the obligation discharged by consensus-leader ordering, batched timestamping, or external clock services), and §6.3 discusses conflict-serializable scheduling.

\subsection{System}

\[K_0 = \varepsilon \qquad T_0 = \varepsilon\]

\[(v_i, K_i, \Xi_i) = \mathrm{execute}(K_{i-1}, \mathrm{tx}_i)\] \[T_i = T_{i-1} \cdot (\mathrm{tx}_i, v_i, \Xi_i) \qquad \text{for } i = 1, 2, \ldots\]

Sequential transaction processing (§6 discusses concurrent scheduling). \(T\) captures the sole non-determinism: external submission order.

\section{Structural Properties}

The six structural properties claimed in §1 follow entirely from the essential structure --- they hold for any atom set \(A\) satisfying the constraints, any set \(B\) of pure built-ins, any program model with send/recall/abort, and any log encoding.

\begin{itemize}
\tightlist
\item
  \textbf{Deterministic replay} --- from append-only \(K\) + deterministic dispatch, given the transaction sequence, re-execution from the empty state reproduces all state.
\item
  \textbf{Immutability of history} --- from append-only logs. No entry is ever modified or deleted.
\item
  \textbf{Authenticated provenance} --- from kernel-controlled caller at position 4. No program can forge or suppress its identity.
\item
  \textbf{Opacity} --- from the absence of any cross-object read mechanism. No object can read another's log or program.
\item
  \textbf{Atomic coordination} --- from transaction semantics. Co-occurrence or mutual destruction of log entries across objects.
\item
  \textbf{History-derived state} --- from log access at position 3. Every object derives its current state from its own complete interaction history.
\end{itemize}

\textbf{Admissibility.} Theorems below assume an instantiation satisfying the §3.2 constraints (partition; \(\mathrm{builtin}\)-on-\(B\); \(\mathrm{alloc}\)-freshness; \(\mathrm{encode}\)-injectivity; program-purity). Theorem 1 additionally requires built-ins, \(\mathrm{encode}\), \(\mathrm{alloc}\), and the interpreter to be deterministic, and program execution to terminate (§3.4).

\textbf{Theorem 1 (Deterministic Replay).} Under the admissibility conditions above, for any transaction sequence \(\mathrm{tx}_1, \ldots, \mathrm{tx}_n\), the state \(K_n\) produced by sequential application of \(\mathrm{execute}\) is uniquely determined.

\subsection{Objects as Information Boundaries}

The persistent object is the only structural boundary in the system. Ephemeral programs inherit context; information flows through them as if they were not there. Built-ins are pure parameterisable functions creating no boundary and maintaining no state. The kernel mediates every crossing of a persistent object's boundary:

\begin{itemize}
\tightlist
\item
  \textbf{Inside \(\to\) outside}: invisible. No program reads another object's log or program, the object registry, or transaction log \(T\). The kernel forbids anonymous send, silent receipt, selective rejection (only transactional abort refuses, destroying the bundle), identity forgery (position 4 is kernel-set), log tampering (position 3 is read-only), and target interception (dispatch is identity-keyed).
\item
  \textbf{Outside \(\to\) inside}: kernel-mediated. Each persistent send: sender chooses target and message; kernel sets caller (position 4) and provides the receiver's history (position 3) and code (position 0); receiver chooses response and may abort. On commit, the kernel records \([\mathrm{caller}, \mathrm{msg}]\) in the receiver's log. External sends are fire-and-forget (discarded on abort); ephemeral sends inherit context; kernel sends produce fresh identities and a birth record.
\end{itemize}

The kernel controls identity allocation, caller assignment, history provision, log writing, dispatch, and transaction atomicity. The sender controls target selection and message content; the receiver controls only the response and whether to abort. The external world controls only transaction submission and input.

\subsection{Identity-as-Capability and Controls}

\textbf{Identity-as-capability.} Knowing an identity is the capability to send. Identities are opaque atoms (the kernel exposes no structure on identities; identities are learned through interaction; allocation pattern-opacity is instantiation-dependent --- see §5.6). Once known, an identity grants unconditional, irrevocable, non-attenuable ability to send.

This forgoes in-kernel attenuation and revocation --- primitives that prior capability systems (Miller\cite{ref9}, Spritely Goblins\cite{ref16}, KeyKOS/EROS\cite{ref27}) treat as core. The design rationale is minimalism: kernel-level revocation would require the kernel to track revocation entitlements and authorise decisions, expanding its surface beyond the send/log/transaction primitives. The kernel instead lets each program specify its own revocation policy through user-level \emph{gating objects} --- opaque persistent intermediaries that inspect caller identity before forwarding.

\textbf{Cooperative fault tolerance.} The receiver's veto is all-or-nothing: it aborts the entire transaction. Programs can implement a ``will-this-succeed'' check by running the same logic and returning yes/no instead of aborting (reliable within a transaction; a check-to-use race across transactions under §6.3). A hypothetical TRY primitive --- catching failures within a transaction --- is deliberately omitted: it would defuse the receiver's only control mechanism and enable traceless probing.

\subsection{History Coupling}

Transactionality couples objects' histories. Without it, each object's log is independently determined. With it, a sender can bundle sends such that entries on different objects' logs atomically co-occur. Three forms:

\textbf{Positive coupling.} The sender bundles sends to \(Q\) and \(R\) in one transaction. \(Q\)'s entry existing implies \(R\)'s entry existing.

\textbf{Negative coupling.} Any receiver can abort, destroying the entire bundle --- including entries on objects it has no direct relationship with. If \(P\) sends to \(Q\) then to \(R\) and \(R\) aborts, \(Q\)'s entry is rolled back.

\textbf{Implicit information channel.} No object can read another's log. But transactional coupling creates a logical entailment: if \(Q\)'s log has entry \(E\), and \(Q\) knows the transaction pattern that produced \(E\), \(Q\) can infer properties of \(R\)'s log without reading it. The co-occurrence guarantee leaks structural information about other objects' histories through the mere existence of one's own entries. Not a direct read --- a deduction from the atomicity contract. Proposition 4 (§4.4) makes this signal precise.

\subsection{Kernel-Level Enforcement}

The properties above admit two readings. A property may hold \emph{for every well-formed program of the system} (kernel-level) or \emph{for some well-formed programs by programmer convention} (user-level, satisfied where programmers choose to).

\textbf{Definition (Kernel-level enforcement).} A property \(P\) is \emph{kernel-level} for system \(S\) iff \(S\) admits no program input whose execution makes \(P\) fail; otherwise \(P\) is \emph{user-level}. For \emph{trace properties} (immutability, atomic commit), this reduces to: the transition function refuses to produce a state in which \(P\) fails. For \emph{hyperproperties} (non-interference, opacity), it requires that executions agreeing on the relevant inputs yield identical post-execution state on objects other than the receiver, identical external sends, and identical response values, excepting the abort-cohort signal of Proposition 4 --- formalised in Theorem 3.

\textbf{Theorem 2 (Read Confinement).} Every read or write of a persistent object's log is mediated by an explicit send to that object via the §3.3 dispatch path. The \(\mathrm{send}_{\mathrm{pers}}\) rule reads only the receiver's \(\mathrm{program}\) and \(\mathrm{log}\) (positions 0 and 3) and appends only to the receiver's log; \(\mathrm{send}_{\mathrm{kern}}\), \(\mathrm{send}_{\mathrm{eph}}\), built-in, external, and special-pair-form dispatches do not reference any persistent object's log. No kernel primitive reads or writes another object's state outside its own dispatch path. \emph{Proof: by inspection of the §3.3 case-split.}

\textbf{Proposition 4 (Abort-cohort signal).} Let \(\mathrm{tx}\) be a transaction whose sub-sends form a cohort \(\{O_1, \ldots, O_k\}\) of receiving persistent objects. The kernel's transaction commit semantics (§3.5) imply that all participants in \(\mathrm{tx}\) --- the initiator and each \(O_i\) --- observe a common commit-or-abort outcome. Consequently, conditional on co-participation, the abort decision of any single \(O_i\) is observable as the abort decision of the bundle. This is the \emph{intentional semantics of atomic coordination} (§4) --- engaging in a multi-party transaction \emph{is} binding one's outcome to the bundle's --- and is not removable without abandoning atomicity. At the in-kernel program level, the signal is visible only to participants in \(\mathrm{tx}\) --- through subsequent inspection of their own logs, which reflect commit (an entry was appended) or abort (no entry was appended). At the system level, \(T\) records the outcome (\(v = \bot\) for receiver-veto aborts), so any party with access to \(T\) observes the bundle's outcome.

\textbf{Theorem 3 (Kernel-level non-leakage modulo abort-cohort signal).} Fix a transaction \(\mathrm{tx}\), a kernel state \(K\), and a persistent object \(O \in \mathrm{Pers}(K)\). Let \(K'\) be any kernel state satisfying \(\mathrm{log}(n, K') = \mathrm{log}(n, K)\) for all \(n \ne O\) and \(\mathrm{program}(O, K') = \mathrm{program}(O, K)\). If during the parallel executions \(\mathrm{execute}(K, \mathrm{tx})\) and \(\mathrm{execute}(K', \mathrm{tx})\) --- (i) every send addressed to \(O\) returns identical response values, and (ii) every send issued by \(O\)'s program is identical between the two runs --- then the two executions agree on the resulting kernel state restricted to objects other than \(O\), the external sends \(\Xi\) initiated by the transaction, and the transaction's response value, modulo Proposition 4's abort-cohort signal when exactly one execution aborts. Information about \(O\)'s log can therefore influence the rest of the system during \(\mathrm{tx}\) only through \(O\)'s responses to its callers, \(O\)'s sends to other persistent objects, \(O\)'s external sends, and the abort-cohort signal.

\emph{Proof sketch.} By structural induction on the sequence of sends issued during \(\mathrm{tx}\). \textbf{Sends to objects \(\ne O\)}: log entries on the target agree by hypothesis on \(K, K'\) together with the inductive hypothesis on \(\Delta\); §3.3 dispatch is deterministic; so response and pending updates agree pointwise. \textbf{Sends to \(O\)}: by (i), \(O\)'s response agrees; by (ii), \(O\)'s onward sends to any target are identical. The post-execution non-\(O\) state, \(\Xi\), and the transaction's response value are functions of these agreed sub-executions, except for the abort-cohort signal of Proposition 4 when one execution aborts and the other commits. \(\square\)

\emph{Remark (on the conditional structure).} Theorem 3 is a \emph{conditional} result: given program-level observability invariance --- (i) and (ii) above --- the kernel propagates that invariance to non-\(O\) state. The kernel's \emph{unconditional} contribution is Theorem 2 (Read Confinement): no primitive reads cross-object state outside dispatch. Program-level non-interference for an opaque \(O\) is therefore reducible to engineering \(O\)'s program to satisfy (i)(ii); the kernel guarantees that satisfaction suffices.

\emph{Corollary (Opacity at read interface).} Information about \(O\)'s log influences the rest of the system only through \(O\)'s deliberate responses, sends, external sends, and the abort-cohort signal. This is the formal content of ``opaque encapsulation'' as claimed in §1.

\textbf{Per-property classification.} All six structural properties claimed in §1 are kernel-level for this kernel under the Definition above. \emph{Trace properties} (transition refuses violating states): immutability of history (no §3 operation removes entries); authenticated provenance (kernel sets caller; programs cannot write position 4); atomic coordination (transaction commits all of \(\Delta\) or none, by §3.5); history-derived state (state is log-projection by §3.2). \emph{Hyperproperties} (executions agreeing on inputs yield identical observations): deterministic replay (Theorem 1); opacity (no cross-object read primitive --- Theorem 3 formalises the non-interference statement). By contrast, the kernel is \emph{not} kernel-level on \emph{resource-bounded execution}: §3.4 assumes termination and defers resource bounding to composition. A non-terminating program halts the kernel; production deployments must compose with deterministic resource metering (gas-style, instruction counts) to preserve Theorem 1's replay/recovery/replication guarantees. Non-deterministic operational fallbacks (wall-clock cut-offs) recover liveness but break determinism.

\section{An Instantiation}

One concrete parameter selection: \(A = \mathbb{N}\), with distinguished elements, six pure built-ins, a five-operation instruction set, and pair-chain log encoding. The instruction set is a \emph{sufficiency demonstration}, not a syntax for direct programmer use; usable surface languages are downstream work (§1).

\subsection{\texorpdfstring{Constraints on \(A = \mathbb{N}\)}{Constraints on A = \textbackslash mathbb\{N\}}}

\begin{itemize}
\tightlist
\item
  \textbf{Decidable equality}: natural number equality.
\item
  \textbf{Distinguished elements}: \(a_{\mathrm{kernel}} = 0\), \(a_{\mathrm{ext}} = 1\), plus program operations and dispatch tags below.
\item
  \textbf{Fresh allocation}: \(\mathrm{alloc}(K \mathbin{+\!\!+} \Delta) = 14 + |\mathrm{log}(0, K \mathbin{+\!\!+} \Delta)|\) --- deterministic, sequential; \(\Delta\) visibility lets multiple creates within one transaction allocate distinct identities.
\end{itemize}

\subsection{Atom Numbering}

Reserved range: \([0, 14)\). Organised in four layers:

\noindent
\small
\renewcommand{\arraystretch}{0.92}
\begin{tabularx}{\columnwidth}{@{} c l X @{}}
\toprule
Number & Name & Role \\
\midrule
\multicolumn{3}{@{}l}{\textbf{Layer 1 --- Identities (fixed)}} \\
0 & KERNEL & System identity --- $a_{\mathrm{kernel}}$, primordial persistent object \\
1 & EXTERNAL & External world --- $a_{\mathrm{ext}}$, sentinel atom; special-cased target \\
\multicolumn{3}{@{}l}{\textbf{Layer 2 --- Instruction-encoding atoms (fixed)}} \\
2 & \texttt{*} & Send \\
3 & \texttt{?} & Recall \\
4 & FAIL & Abort \\
5 & \texttt{!} & Quote \\
\multicolumn{3}{@{}l}{\textbf{Layer 3 --- Dispatch tags (fixed)}} \\
6 & PAIR\_TAG & Pair closure tag \\
7 & EXTERNAL\_TAG & External target tag \\
\multicolumn{3}{@{}l}{\textbf{Layer 4 --- Built-ins (parameterisable; §5.3)}} \\
\multicolumn{3}{@{}l}{\quad 8 HEAD, 9 TAIL, 10 PAIR,} \\
\multicolumn{3}{@{}l}{\quad 11 EQUAL, 12 BRANCH, 13 INCREMENT} \\
\bottomrule
\end{tabularx}
\renewcommand{\arraystretch}{1.0}

Fixed atoms \([0, 8)\) define system structure. Built-in atoms \([8, 14)\) are replaceable with any set satisfying the built-in constraints.

\subsection{\texorpdfstring{Built-in Objects \(B\)}{Built-in Objects B}}

Signature: \(\mathrm{builtin} : A \times \mathcal{S} \to \mathcal{S} + \{\bot\}\). Six built-in objects, all pure. Booleans use the structural atom/pair distinction: atom = true, pair = false.

{\small
\renewcommand{\arraystretch}{0.85}
\setlength{\abovedisplayskip}{-2pt}
\setlength{\belowdisplayskip}{2pt}
\[
\mathrm{builtin}(n, m) = \begin{cases}
a & n = 8,\; m = [a, b] \quad \text{(HEAD)} \\
b & n = 9,\; m = [a, b] \quad \text{(TAIL)} \\
{[6, m]} & n = 10 \quad \text{(PAIR)} \\
0 & n = 11,\; m = [a, b],\; a = b \quad \text{(EQUAL)} \\
{[a, b]} & n = 11,\; m = [a, b],\; a \neq b \\
x & n = 12,\; m = [t, [x, [y, \_]]],\; t \in A \\
y & n = 12,\; m = [t, [x, [y, \_]]],\; t \in \mathcal{S} \times \mathcal{S} \\
m + 1 & n = 13,\; m \in A \quad \text{(INCREMENT)} \\
\bot & \text{otherwise}
\end{cases}
\]
}

PAIR returns a closure \([6, a]\). The instantiation-defined pair-form dispatch \(\Pi\) (§3.2) is given by \(\Pi([6, a], m) = [a, m]\); \(\Pi(t, m) = \bot\) otherwise. The remaining pair classification: \(\mathrm{Ext} = \{[7, x] \mid x \in \mathcal{S}\}\) (pairs headed by EXTERNAL\_TAG); \(\mathrm{Eph} = (\mathcal{S} \times \mathcal{S}) \setminus (\mathrm{dom}(\Pi) \cup \mathrm{Ext})\). Pair closures thus dispatch as pure construction with no effects, with priority over the generic ephemeral case. Pair closures are structurally indistinguishable from any pair whose head is PAIR\_TAG (6); programs can construct closures directly. BRANCH is the sole discrimination primitive (atom-vs-pair), replacing pair-testing and conditional selection. INCREMENT is the sole arithmetic primitive (specific to \(A = \mathbb{N}\)). Together with send and recursion via recall of position 0, these built-ins suffice for general computation over \(\mathcal{S}\) --- Church-encoded naturals are one standard encoding; the atom/pair distinction with BRANCH is another.

\subsection{Instruction Set}

Five operations: push, quote (\(!\)), send (\(*\)), recall (\(?\)), abort (FAIL). A program is an s-expression interpreted as a linked list.

\noindent
\small
\begin{tabularx}{\columnwidth}{@{} l X @{}}
\toprule
Instruction & Action \\
\midrule
$4$ (FAIL) & Abort: produce $\bot$ \\
$n \in A \setminus \{4\}$ & Return $L[|L|-1]$ (top of stack) \\
$[2, k]$ (send) & Send $L[|L|-2]$ to $L[|L|-1]$; push result; continue with $k$ \\
$[3, [j, k]]$ (recall, $j \in A$, $j < |L|$) & Push $L[j]$; continue with $k$ \\
$[5, [d, k]]$ (quote) & Push $d$ literally (never interpreted); continue with $k$ \\
$[d, k]$ (push) & Push $d$; continue with $k$ \\
\bottomrule
\end{tabularx}

Execution begins with \(\mathrm{run}(L_0, p)\) where \(L_0 = [p, m, s, l, c]\) binds program, message, self, log, caller. This instantiates §3.4's signature: the context list is the initial stack (grown with intermediate values on recursion), and the program is the initial instruction (replaced by the continuation \(k\) on recursion). Every program can recall its context (positions 0--4); recall is the kernel's \emph{intra-object} read primitive --- programs cannot read any other object's program or log. Inter-object observation is necessarily via send, which produces a log entry; the absence of non-mutating cross-object reads is load-bearing for audit (every cross-object observation is auditable).

\needspace{12\baselineskip}

The full interpreter:

\setlength{\abovedisplayskip}{-2pt}
\begin{align*}
\mathrm{run}&(L, i, K, \Delta, \Xi) = \\
&\begin{cases}
\bot & i = 4 \\
(L[|L|-1],\; \Delta,\; \Xi) & i \in A \setminus \{4\} \\
\bot \text{ if } v {=} \bot, \text{else } \mathrm{run}(L \cdot v, k, K, \Delta', \Xi') & i = [2, k] \;\; (\dagger) \\
\mathrm{run}(L \cdot L[j], k, K, \Delta, \Xi) & i = [3, [j, k]],\; j < |L| \\
\bot & i = [3, \_] \text{ otherwise} \\
\mathrm{run}(L \cdot d,\; k,\; K,\; \Delta,\; \Xi) & i = [5, [d, k]] \\
\bot & i = [5, \_] \text{ otherwise} \\
\mathrm{run}(L \cdot d,\; k,\; K,\; \Delta,\; \Xi) & i = [d, k]
\end{cases}
\end{align*}

where \((\dagger)\) abbreviates: \[(v, \Delta', \Xi') = \mathrm{send}(L[|L|-1],\, L[|L|-2],\, L[2],\, L[3],\, L[4],\, K, \Delta, \Xi)\]

The execution log is scratch --- always discarded after execution completes. Context transitions: send to a persistent object switches self, log, caller; send to an ephemeral program inherits them unchanged. Position 3 provides read-your-writes within a transaction.

\subsection{Pair-Chain Encoding}

The log encoding \(\mathrm{encode} = \lceil\cdot\rceil\) represents logs as pair-chains terminated by 0:

\[\lceil\varepsilon\rceil = 0\] \[\lceil(a, b) \mathbin{.} t\rceil = [[a, b], \lceil t \rceil]\]

A log with entries \(e_0, e_1, e_2\) is \([e_0, [e_1, [e_2, 0]]]\). The terminator 0 coincides with the kernel identity (\(a_{\mathrm{kernel}}\)); these roles do not conflict --- 0 as terminator appears in a structural position, while 0 as kernel identity appears in semantic contexts.

\subsection{Identity Allocation}

Sequential: \(\mathrm{id} = 14 + |\mathrm{log}(0, K \mathbin{+\!\!+} \Delta)|\). Each object creation increments the count of entries on the kernel's object registry. Deterministic.

The §3 essential semantics leaves \(\mathrm{alloc}\) abstract. The sequential counter here is \emph{pattern-opaque only by programmer convention} --- under the §4.4 demarcation criterion, identity-pattern opacity is \emph{user-level} (a program can compute INCREMENT on a known identity to probe for adjacent allocations). Alternative instantiations (e.g., \(\mathrm{alloc}(K) = h(\mathrm{nonce}(K))\) for a one-way hash \(h\)) preserve kernel-level pattern-opacity. We chose sequential for clarity; the §3-level claims are independent of this choice.

\textbf{Sufficiency, not expressibility.} The instruction set demonstrates that general computation is encodable but at significant syntactic cost: conditional dispatch via BRANCH requires constructing nested pair arguments (\textasciitilde18 instructions per branch point). A practical instantiation would provide richer control-flow primitives; the minimal set here establishes a lower bound on what suffices.

\section{Composition}

\subsection{Locality}

The kernel's per-object semantics imply that it can run on any subset of object state without affecting its structural guarantees. Two equivalent formulations:

\textbf{Partial-knowledge execution.} A \emph{local} kernel runs with state \(K_{\mathrm{local}} \subset K_{\mathrm{global}}\) --- containing only some objects' histories --- and executes transactions that touch only those objects. \(\mathrm{send}(t, m, \ldots)\) looks up \(\mathrm{program}(t, K \mathbin{+\!\!+} \Delta)\) and \(\mathrm{log}(t, K \mathbin{+\!\!+} \Delta)\); if \(t\) is in scope these are defined and the transaction proceeds, otherwise \(t\) falls through to §3.3's catch-all and the send returns \(\bot\). Use cases: sharding (§6.2, each shard's partition as \(K_{\mathrm{local}}\)), privacy (principals verify only their own objects' interactions), independent audit (replay dispute-relevant interactions without consulting unrelated state).

\textbf{Compositionality.} Equivalently, multiple kernel instances \(\mathcal{K}_1, \ldots, \mathcal{K}_n\) with pairwise-disjoint persistent object identities can run independently. Cross-instance interactions are mediated by external sends --- an external send from \(\mathcal{K}_i\) with destination in \(\mathcal{K}_j\) becomes a new transaction submitted to \(\mathcal{K}_j\) via \(a_{\mathrm{ext}}\). Each \(\mathcal{K}_i\) independently maintains the six structural properties of §3 within its own scope (no operation references another instance's state; Theorems 1 and 2 apply unchanged within each instance).

\textbf{Boundary degradation.} Authenticated provenance and atomicity do not extend across scope boundaries by default. A receiver of a cross-scope interaction sees the kernel-assigned caller \(a_{\mathrm{ext}}\), not the originating object; and the originating transaction commits independently of consequent actions in other scopes. Recovering cross-scope provenance or atomicity requires additional protocol mechanisms.

\subsection{Composition Modes}

\textbf{Replication and consensus.} Deterministic replay (Theorem 1) is the state-machine replication property\cite{ref23}: identical transaction sequences yield identical \(K\). The transaction log \(T\) is the command log a consensus or replication protocol (Raft, Paxos) agrees upon and disseminates; replicas --- including mutually-distrusting parties verifying by local re-execution --- independently apply transactions in the agreed order and reach identical state. Wrapping the kernel requires no kernel modification, only that the protocol agree on and deliver \(T\).

\textbf{Sharding.} A kernel's state \(K\) is a flat sequence of per-object log entries; nothing in the formal semantics requires global \(K\) to be physically co-located. The object space can be partitioned across shards, each shard's kernel maintaining \(K\) entries only for objects in its partition. Per-shard transactions are atomic by the kernel's standard semantics; cross-shard interactions become external sends (\(a_{\mathrm{ext}}\) destinations routed to other shards) wired up by the deployment layer. Cross-shard atomicity, when required, needs additional protocol layered over the kernel (two-phase commit orchestrated by a coordinator; consensus agreeing on a total order before per-shard execution) --- not addressed by the kernel.

\subsection{Conflict-Serializability and Per-Property Survival}

The sequential transaction processing of §3 is sufficient but not necessary: all six properties survive under any conflict-serializable scheduling, allowing replication and concurrent execution to reorder non-conflicting transactions.

\textbf{Conflict, commutativity, and the parallelism floor.} Two transactions \emph{conflict} if they append log entries to at least one common persistent object. Since every cross-\emph{persistent}-object access appends to the receiver's log (built-ins, ephemerals, pair-forms, externals do not), conflicts are \emph{write-write only} and the conflict graph is dense --- any pair of transactions touching a common persistent object conflicts. A schedule is \emph{conflict-serializable}\cite{ref17} if equivalent in effect on \(K\) to some serial order.

This dense conflict definition has a non-trivial parallelism cost: any persistent object touched by multiple transactions linearises all such transactions, even when they concern logically-independent state. Single-hot-object workloads serialise (as in Calvin\cite{ref25}); shared coordinator/registry objects become bottlenecks. The kernel does not capture the read-write parallelism that conventional optimistic concurrency control (SSI) extracts via separate read/write tracking --- the trade-off is for structural opacity, in-state provenance, and uniform deterministic replay. \emph{Non-mutating sends} would be a separate design extension, not currently in the kernel's structural model. Specific scheduling strategies are deployment concerns.

\textbf{Per-property survival.} Each property survives by a per-property argument:

\begin{itemize}
\tightlist
\item
  \emph{Immutability} --- the log remains append-only regardless of schedule.
\item
  \emph{Authenticated provenance} --- each entry records the kernel-assigned caller, set at context switch independently of global order.
\item
  \emph{Opacity} --- no send case exposes another object's log or program, regardless of concurrent scheduling.
\item
  \emph{History-derived state} --- under conflict-serializable scheduling, the log entries visible to an object are those committed by transactions serialised before the current one, plus pending entries from earlier completed invocations within the current transaction (read-your-writes via \(\Delta\), §3.2).
\item
  \emph{Atomic coordination} --- per-transaction: each transaction either commits all entries or none, independent of others.
\item
  \emph{Deterministic replay} --- replaying transactions in the order recorded in \(T\) reproduces \(K\); for conflict-free transactions, any equivalent interleaving produces the same result.
\end{itemize}

\textbf{Non-determinism and identity allocation.} Non-determinism now derives from submission order plus serialisation choices, both captured in \(T\). Conflict-aborts produce no \(T\) entry (pre-serialisation scheduling decisions); only receiver-veto aborts (programmatic FAIL) appear in \(T\) with \(v = \bot\). The conflict set is dynamic (computed during execution); Calvin-style reconnaissance or optimistic retry handle prediction-misses. Receiver-veto aborts depend on the visible commit prefix under the chosen serial order, so Theorem 1's replay reproduces the abort pattern. \(T\) is the post-serialisation log. A concurrent instantiation uses a partitioned or hash-based identity allocation rather than the §5 counter.

\subsection{Log Canonicality and Retention}

The persistent state \(K\) is canonical and non-truncatable: truncation breaks deterministic replay (Theorem 1) and the audit guarantee on which authenticated provenance rests. Derived state --- annotations, caches, materialised views --- may be reclaimed; \(K\) may not. \(|K|\) is therefore monotonically growing.

\textbf{State annotations (compiler transformation).} Naive per-message cost is \(O(|\mathrm{history}|)\) --- \(O(N^2)\) over \(N\) messages to one object. For programs \(P\) admitting a fold structure \((\sigma_0, \mathrm{step}, \mathrm{use})\) with \(P(H, \mathrm{msg}) = \mathrm{use}(\mathrm{foldl}(\mathrm{step}, \sigma_0, H), \mathrm{msg})\), a source-to-kernel compiler can emit \(P'\) that appends a checkpoint self-send carrying \(\sigma\) after each invocation; subsequent calls read the latest checkpoint from position 3 --- \(O(1)\) amortised per call (one extra log entry per invocation). The base kernel of §3 is unchanged; \(P'\) is observationally equivalent to \(P\) on responses, and all six properties survive. Most common programs (counters, balances, key-value stores, ACLs, FSMs) admit this form. Under conflict-serializable schedules (§6.3), non-idempotent folds need additional care.

\textbf{Crash recovery.} The kernel's persistent state is \((K, T)\), plus any commit-time \(\Delta_i\) pending merge into \(K\) (used for crash recovery, see below); the kernel does not carry any other state across transactions. \(T\) is the \emph{canonical} log; \(K\) is its deterministic materialisation under Theorem 1. The pending buffer \(\Delta\) is volatile; mid-transaction crash discards \(\Delta\) and the transaction is treated as never started (atomicity preserved by §3.5). On commit, \((T_i, \Delta_i)\) are persisted atomically as a single durability unit via group commit, alongside the resulting length \(|K_i|\). On restart, recovery walks \(T\) from the last durable position; for each \(T\)-entry not yet reflected in \(K\) (comparing \(|K|\) to the recorded \(|K_i|\)), the corresponding durable \(\Delta_i\) is appended to \(K\). There is no undo, and no re-execution: the durable \(\Delta_i\)'s suffice for \(K\)-extension. External sends \(\Xi_i\) are dispatched at-least-once on a read-side traversal of \(T\) after \(T_i\) is durable; external endpoints must deduplicate on \(\mathrm{tx}_i\). The underlying durability layer is a deployment concern. Byzantine kernel-implementation faults are out of scope; cross-replica \(K\)-divergence is detected by Theorem 1's replay verification.

\section{Patterns}

Common patterns encodable on the kernel. By \emph{encodable} we mean: for each pattern, there exists a set of persistent objects with specific programs that produces the desired observable behavior (responses and external sends) for message sequences within the pattern's defined interface. \textbf{What is not encodable.} Operations requiring direct write-access to another object's program or log are structurally precluded by the kernel's no-cross-object-write property: Smalltalk's \texttt{instVarAt:put:} (writing another object's instance variables), Self's \texttt{become:} (atomic identity swap), and first-class continuations that capture another object's execution stack cannot be encoded --- they would violate Theorem 2 (Read Confinement).

\subsection{Object-System Encodings}

\textbf{Self-style delegation.} A program obtains a parent's code (via literal or message) and sends to it via ephemeral dispatch, which inherits self/log/caller. The parent's code thus runs \emph{as the child}: position 2 returns the child's identity, position 3 the child's log, position 4 the child's caller, position 0 the parent's currently-executing code. A Self object with slots \(\{a{:}\,1, b{:}\,2, \text{parent}{:}\,P\}\) encodes as a persistent object pattern-matching on the message (slot name): on hit return the value; on miss ephemeral-send to \(P\)'s code. Mirrors map to self-reads of position 3; assignment slots use the code-change pattern; multiple parents are sequential delegation attempts.

\textbf{Smalltalk-style classes} add one indirection over Self-style delegation: method resolution goes through a class object (persistent send), then the resolved method runs with the instance's context (ephemeral send). Classes accumulate method definitions in their logs (\([\mathrm{DEFINE}, \mathrm{selector}, \mathrm{code}]\)) with the current dictionary as a state annotation (§6); they respond to \([\mathrm{LOOKUP}, \mathrm{selector}]\) (dictionary check, recurse to superclass on miss) and \([\mathrm{NEW}, \mathrm{state}]\) (create instance referencing class identity). Instances persistent-send \([\mathrm{LOOKUP}, \mathrm{selector}]\) then ephemeral-send \(\mathrm{args}\) to the returned method --- exactly Smalltalk's method execution semantics; instance variables are the state annotation (position 3). \texttt{super} sends \([\mathrm{LOOKUP}, \mathrm{selector}]\) to the defining class's superclass; metaclass towers terminate at a self-handling root; \texttt{doesNotUnderstand:} is returned when lookup reaches the root without a match. The class can amortize lookup by caching a compiled dispatcher as its state annotation.

\textbf{Meta-object protocols.} Kernel dispatch is fixed (the six send cases), but within persistent objects, programs have full control over method dispatch, delegation, and intercession --- the program \emph{is} the protocol. This differs from CLOS's MOP\cite{ref10} in that there is no fixed method-combination or slot-access protocol to extend. Conventional MOP messages can be standardised as a library convention (\([\mathrm{LOOKUP}, \mathrm{selector}]\) returns method or delegates; \([\mathrm{INVOKE}, \mathrm{selector}, \mathrm{args}]\) looks up and sends; \([\mathrm{DELEGATE}, \mathrm{parent}, \mathrm{msg}]\) inherits context; \([\mathrm{SLOTS}]\) enumerates) --- opt-in interoperation, not a kernel extension.

\subsection{Pattern Catalog}

Most common object-system patterns (self-recursion, proxies, conditional dispatch, pub/sub, state machines) follow directly from recall + send + BRANCH. A few are worth highlighting:

\begin{itemize}
\tightlist
\item
  \textbf{Clone / code change.} Recall position 0 + send to kernel produces a same-program new object; transforming the recalled program before sending implements modification. Code change is the \emph{bootloader pattern}: a fixed program traverses its log for the latest deposited code entry and dispatches to it.
\item
  \textbf{Access control via identity gating.} A gating object inspects caller against a predicate before returning \(Q\)'s identity --- the sole point of control, since identity-as-capability is irrevocable once distributed.
\item
  \textbf{Asynchronous messaging.} A program sends to \([7, \mathrm{addr}]\) (the EXTERNAL\_TAG pair); the external world re-submits as a new transaction. \emph{Auth caveat:} the re-submitted transaction has caller \(= a_{\mathrm{ext}}\), breaking original-sender authentication (same phenomenon as §6.1 boundary degradation).
\item
  \textbf{Phase-gated state machines.} A persistent object branches on its log structure (e.g., \(\mathrm{TAIL}(L[3])\) atom vs.~pair) to determine its current protocol phase; per-phase EQUAL against \(L[4]\) enforces caller-gating. A two-party escrow (buyer deposits in phase 0, seller confirms in phase 1, query returns ``complete'' in phase 2) falls out directly --- wrong-phase callers abort by failed EQUAL, the escrow's state is its log length.
\end{itemize}

\subsection{Case Studies}

The two case studies below serve different purposes. \textbf{Case Study 1 (Delegation Chain with Accountability) is the central mechanism demonstration}: it isolates the ephemeral/persistent classification step as the sole load-bearing difference between two otherwise-identical traces, showing that accountability (which principal is recorded in the receiver's log) follows directly from target shape. \textbf{Case Study 2 (Sealed-Bid Auction) is integrative}: it shows how multiple structural properties combine to serve a specific multi-party coordination pattern. The argument is not that all six properties are jointly required for any single case study, but that the six are \emph{individually motivated} by these patterns and \emph{jointly compatible} under one kernel design. Case Study 2 relies on the operator-faithful trust regime; its EVM contrast is scope-conditional. Bidder proxies represent external principals; proxy-ownership is enforced by the wrapping protocol (§1), not the kernel.

\subsubsection{Case Study 1: Delegation Chain with Accountability}

\textbf{Setup.} A server B (identity 15) that returns \([\mathrm{caller}, \mathrm{msg}]\), used to observe which identity the server perceives as its caller. Three experiments:

\begin{enumerate}
\def\labelenumi{\arabic{enumi}.}
\tightlist
\item
  \textbf{Ephemeral delegate P.} Object A (14) delegates to ephemeral code P, which sends to B. Since P is ephemeral, it inherits A's execution context --- B sees caller \(= 14\) (A).
\item
  \textbf{Different ephemeral delegate P'.} Object A' (16) uses delegate P' (which wraps the message as \([42, \mathrm{msg}]\)). B still sees caller \(= 16\) (A'), because the ephemeral context is still A'.
\item
  \textbf{Persistent delegate C.} Object A'\,' (18) sends to persistent object C (17), which forwards to B. B sees caller \(= 17\) (C), not \(18\) (A'\,').
\end{enumerate}

\textbf{Transaction trace.}

\noindent
\small
\begin{tabularx}{\columnwidth}{@{} c X l c @{}}
\toprule
Tx & Action & Result & B sees caller \\
\midrule
1 & Create A(14), B(15) & $15$ & --- \\
2 & Send 100 to A $\to$ P $\to$ B & $[14, 100]$ & A (14) \\
3 & Create A'(16) & $16$ & --- \\
4 & Send 200 to A' $\to$ P' $\to$ B & $[16, [42, 200]]$ & A' (16) \\
5 & Create C(17), A''(18) & $18$ & --- \\
6 & Send 300 to A'' $\to$ C $\to$ B & $[17, 300]$ & C (17) \\
\bottomrule
\end{tabularx}
\normalsize

\textbf{Key insight.} The \emph{only} difference between transactions 2 and 6 is whether the target is classified as ephemeral (pair target → context inherited) or persistent (atom target with birth record → context switched). Everything downstream --- provenance, opacity, accountability --- flows from this single classification step. Ephemeral delegation preserves the delegator's identity (accountability without indirection), while persistent delegation introduces a new principal (the delegate acts under its own identity). This is the kernel's central structural distinction, and it emerges from the same send operation applied to different target shapes.

\subsubsection{Case Study 2: Sealed-Bid Auction}

\textbf{Setup.} A single kernel operator (a regulated auctioneer, service provider, or trusted intermediary). Each bidder is represented in-kernel by a persistent proxy object \(\beta_i\); bids must be sealed against each other while the operator is trusted not to pre-reveal. (Operator-level mutual distrust requires additional layering --- see closing remarks.) Participants: bidders \(\beta_1, \ldots, \beta_n\) and an auctioneer object \(A\). \emph{The auctioneer constructs each bid-object} --- \(A\) creates \(B_i\) rather than the bidder, so \(A\) knows by construction that \(B_i\) runs the canonical program \(p_B\) (otherwise a bidder could substitute a \(B_i\) that returns an arbitrary value on REVEAL). Each bidder \(i\) (a) sends \([\mathrm{REGISTER}]\) to \(A\); (b) \(A\) creates \(B_i\) with program \(p_B\), sends \([\mathrm{INIT}, \beta_i]\) to \(B_i\) (binding \(\beta_i\) as the authorised storer), self-sends \([\mathrm{TRACK}, \beta_i, B_i]\) (recording the binding in \(A\)'s own log), and returns \(B_i\) to \(\beta_i\); (c) \(\beta_i\) sends \([\mathrm{STORE}, \mathrm{bid}]\) to \(B_i\).

\textbf{Programs.} \(A\)'s program: on \([\mathrm{REGISTER}]\) from caller \(\beta\), kernel-create \(B_i\) with \(p_B\), send \([\mathrm{INIT}, \beta]\) to \(B_i\), self-send \([\mathrm{TRACK}, \beta, B_i]\), return \(B_i\); on \([\mathrm{REVEAL\_ALL}]\), walk own log for TRACK entries, send REVEAL to each \(B_i\), return the bidder with the highest revealed bid. \(B_i\)'s canonical program \(p_B\): recover \(A\) from its birth-record caller (the kernel-set caller of the first log entry) and \(\beta\) from the INIT entry; on \([\mathrm{STORE}, \mathrm{bid}]\) from caller \(= \beta\), the kernel auto-logs the entry; on \([\mathrm{REVEAL}]\) from caller \(= A\), return the stored bid; otherwise FAIL.

\textbf{Transaction trace.}

\noindent
\small
\begin{tabularx}{\columnwidth}{@{} c X X @{}}
\toprule
Tx & Action & Result \\
\midrule
1 & $\beta_1$ sends $[\mathrm{REGISTER}]$ to $A$ & $A$ creates $B_1$, INIT $\beta_1$, TRACK; returns $B_1$ \\
2 & $\beta_1$ sends $[\mathrm{STORE}, 100]$ to $B_1$ & $B_1$'s log: $[\beta_1, [\mathrm{STORE}, 100]]$ \\
3 & $\beta_2$ sends $[\mathrm{REGISTER}]$ to $A$ & $A$ creates $B_2$, INIT $\beta_2$, TRACK; returns $B_2$ \\
4 & $\beta_2$ sends $[\mathrm{STORE}, 150]$ to $B_2$ & $B_2$'s log: $[\beta_2, [\mathrm{STORE}, 150]]$ \\
5 & REVEAL\_ALL to $A$ & atomically: $A \to B_1$ (returns 100), $A \to B_2$ (returns 150); $A$ returns $\beta_2$ \\
\bottomrule
\end{tabularx}
\normalsize

In Tx 5, both REVEAL sends are sub-sends of one transaction. If any \(B_i\) aborts (returns \(\bot\)), the whole transaction aborts --- selective withdrawal after observing others' bids is structurally impossible.

\textbf{Property analysis.} Each of the six properties is load-bearing:

\begin{itemize}
\tightlist
\item
  \emph{Opacity}: \(B_i\)'s log is invisible to other bidder-programs; bids are sealed without hash commitments or zk-proofs. Sealing is structural rather than cryptographic.
\item
  \emph{Authenticated provenance}: \(B_i\)'s birth record (kernel-set caller \(= A\)) identifies its creator; the INIT entry binds \(\beta_i\); STORE and REVEAL callers are verified against these kernel-set identities (no program can write position 4). \(A\)'s construction of \(B_i\) closes the trust chain --- the canonical program is what \(A\) supplied.
\item
  \emph{Atomic coordination}: reveal across all \(B_i\) occurs in one transaction; no bidder can selectively withhold after observing others.
\item
  \emph{Immutability of history}: once \([\beta_i, [\mathrm{STORE}, \mathrm{bid}]]\) is recorded in \(B_i\)'s log, the bid cannot be revised; commitment is irreversible.
\item
  \emph{History-derived state}: \(A\)'s ``committed bidders'' set and the winner are both folds over \(A\)'s log. No mutable auctioneer state.
\item
  \emph{Deterministic replay}: any party with appropriate access can re-execute over \(T\) to verify the winner.
\end{itemize}

\textbf{Operator-level observation.} Within this trust model, the kernel operator can read \(K\) --- the structural opacity is against \emph{bidder-programs}, not against the operator. Nevertheless, the operator cannot fabricate log entries: any divergence between the operator's claimed \(K\) and the deterministic replay of \(T\) is externally detectable. The kernel separates operator \emph{read} access (unprotected) from operator \emph{write} access (constrained by replay-verifiability) --- providing an ``honest operator under audit'' guarantee even without cryptography.

\textbf{Contrast with existing systems.} EVM-based sealed-bid auctions require explicit hash-commit / reveal protocol code with gas cost at every step; the kernel does this structurally (under stated operator trust). Akka Classic's \texttt{sender} is settable by the caller (\texttt{tell(msg,\ anyRef)}); Akka Typed removes implicit sender entirely, making sender identity a protocol convention. Either way, authentication is at the protocol layer (typically via signed messages) rather than kernel-enforced --- \(B_i\) cannot rely on a structural platform guarantee that REVEAL came from \(A\). Datomic exposes facts to readers, violating opacity. Each existing system fails at least one of the simultaneous properties this protocol relies on.

\textbf{Compilation to a multi-party protocol (Viaduct sketch).} When no party can be designated trusted operator, the kernel composes with a \emph{cryptographic compiler} in the sense of Viaduct\cite{ref24}: kernel programs become specifications compiled into multi-party protocols using MPC, commitments, and ZK proofs. For kernel programs over a fixed set of principal-identities (known at compile time), object-identity structure determines the label lattice: each persistent object's identity \emph{is} its confidentiality label, and send dispatch induces label flows (a value sent from \(A\) to \(B\) acquires label \(\{A\}\) at \(B\)). The compiler identifies cross-label flows and inserts MPC for cross-bidder aggregation, ZK for winner-announcement declassification, and per-bidder encryption against operator read access. Obligations: (a) aborts must be uncorrelated with private values (compiler-checkable IFC; otherwise abort patterns leak distribution); (b) the auctioneer program must be a commutative function of bids --- a program-level property the developer ensures, with §3.2 purity already excluding operator-private state. The kernel's dynamic-allocation capability extends beyond Viaduct's static-label framework; this sketch covers only the static-allocation subset, with dynamic extension as future work.

\section{Related Work}

The kernel realises the six structural properties of §1 as kernel-level guarantees (opacity against program-level participants under operator-faithful trust; atomic commit per kernel instance). It deliberately omits attenuable capabilities, resource bounding, read-only reads, and async/promise composition (§4 gating, §3.4 deferred resource model, §6.3 read-set-as-write-set, §7 fire-and-forget external sends). The closest existing systems share subsets: Akka Persistence (immutable history, history-derived state, partial opacity); Goblins (opacity, local atomic commit); Datomic (immutable history, history-derived state, atomic commit). Detailed divergences below, organised by tradition.

The comparison below summarises each system's property coverage under §4.4's Definition of kernel-level enforcement; cell verdicts trace to the per-system prose in §§8.1--8.3.

\noindent
\footnotesize
\setlength{\tabcolsep}{4pt}
\begin{tabular}{@{} l c c c c c c @{}}
\toprule
System & Prov & Opac & Atom & Repl & Imm & Hist \\
\midrule
Smalltalk\textsuperscript{14,15}     & U   & U   & ---  & ---  & ---  & ---  \\
Self\textsuperscript{1,2}            & U   & U   & ---  & ---  & ---  & ---  \\
EVM\textsuperscript{5}               & U   & K*  & K    & K    & K    & U    \\
Akka Persist.\textsuperscript{22}    & U   & U   & ---  & U    & K    & K    \\
Datomic\textsuperscript{18,19}       & --- & --- & K    & K    & K    & K    \\
Clinger\textsuperscript{20}          & U   & K   & ---  & K    & K    & K    \\
Spritely Goblins\textsuperscript{16} & U   & K   & K    & K    & K    & U    \\
Nock/Urbit\textsuperscript{3,4}      & U   & K   & U    & K    & K    & K    \\
\textbf{Send kernel}                 & \textbf{K} & \textbf{K} & \textbf{K} & \textbf{K} & \textbf{K} & \textbf{K} \\
\bottomrule
\end{tabular}\\[2pt]
\textit{Headers:} Prov = auth. provenance; Opac = opacity; Atom = atomic commit; Repl = det. replay; Imm = immut. history; Hist = hist-derived state. \textit{Cells:} K = kernel-level (transition refuses violating states / structural property of the dispatch); U = user-level (held by programmer convention or per-application implementation); --- = system lacks the concept or property is not applicable. * EVM storage is per-contract opaque under the EVM opcode set; the underlying state trie is publicly readable from outside the VM, so opacity is conditional on the deployment exposing only the VM-level interface.
\normalsize

The kernel row is the unique fully-kernel-level coverage of these six properties \emph{under the Definition}; alternative designs distribute the kernel/user-level boundary differently (e.g., EVM accepts user-level provenance via \texttt{DELEGATECALL} to make composition more permissive; Datomic accepts no caller concept because facts are unauthenticated).

\subsection{Object Systems and Messaging}

\textbf{Smalltalk} (Kay 1993\cite{ref14}, Ingalls 1981\cite{ref15}). Shared: objects-and-messaging foundation, opaque references, receiver controls response. Divergences: assignment is not a message (kernel: send is the only cross-object interface); mutable instance variables vs.~append-only history; class-based vs.~flat dispatch; \texttt{instVarAt:}/\texttt{become:} break encapsulation (kernel: structurally precluded).

\textbf{Self} (Ungar and Smith 1987\cite{ref1}, Chambers et al.~1989\cite{ref2}). Shared: prototype orientation, delegation via message send, no class/instance distinction. Divergences: mutable slots vs.~append-only history; mirrors expose state externally vs.~structural opacity; explicit parent slots vs.~ephemeral-send context inheritance; no transactional coupling. Self-style prototypes and delegation are encodable here (§7); operations requiring direct write-access to another object (Self's \texttt{become:}) are structurally precluded by Theorem 2.

\textbf{Capability-based security} (Dennis and Van Horn 1966\cite{ref8}, Miller 2006\cite{ref9}, Miller et al.~2003\cite{ref21}). Shared: identity-as-capability, unforgeable references. Divergences: capability systems typically support attenuation/revocation (kernel: non-attenuable, irrevocable); the kernel additionally couples capability use to a per-object append-only audit log.

\textbf{EVM} (Wood 2014\cite{ref5}). Shared: transaction-based state transitions, deterministic execution. Divergences: mutable storage slots vs.~append-only logs; code-data separation vs.~code-is-data; transparent contracts (code readable on-chain) vs.~opaque objects; explicit gas metering vs.~kernel-level none (composition-layer scope); ad-hoc \texttt{DELEGATECALL} vs.~structural delegation via context inheritance.

\subsection{Actors and Process Calculi}

\textbf{Actor model} (Hewitt et al.~1973\cite{ref6}, Agha 1986\cite{ref7}, Clinger 1981\cite{ref20}). Shared: message passing, encapsulation, objects-as-principals. Clinger's histories are the direct intellectual ancestor of ``state is history''; the distinction is denotational (Clinger) vs.~operational-runtime (kernel) --- Clinger's histories are system-wide and external; the kernel's logs are per-object and opaque. Further divergences: actors evolve via Agha's \texttt{become} (behavior replacement; kernel: append-only history retains all prior behavior); no transactional coupling (kernel: atomic multi-send bundles); no mandatory caller disclosure (kernel: position 4).

\textbf{Pi-calculus} (Milner 1999\cite{ref11}, Sangiorgi and Walker 2001\cite{ref12}). Shared: name-passing as structuring principle (identity distribution = name mobility; \(\nu\) = fresh allocation; interaction mediated by names). Divergences: pi-calculus has no persistence (kernel: logs and transactions); replication primitive in pi-calculus, encodable here; processes anonymous vs.~kernel-assigned persistent identity. The kernel is a synchronous name-passing calculus augmented with persistent named processes, append-only observation, and atomic transactions.

\textbf{Erlang/OTP} (Armstrong 2003\cite{ref13}). Shared: per-process sequential message handling, process isolation, hot code loading (code-change pattern), fresh identity on creation. Divergences: ephemeral current state via tail-recursive evolution over immutable data vs.~retained append-only history; no authenticated caller identity (kernel: kernel-enforced); no transactional coupling vs.~atomic multi-send bundles; selective receive vs.~submission-order processing.

\textbf{Spritely Goblins} (Lemmer-Webber et al.~2023\cite{ref16}). Closest existing system combining transactions with capability security. Opacity is kernel-level via vat-bounded isolation (no cross-vat read primitive); atomic commit is kernel-level via transactional rollback over actormap snapshots; deterministic replay is kernel-level via deterministic vat execution; immutable history is kernel-level via generational transactormap chains, where each committed generation is stable and supports time-travel debugging. Authenticated provenance is user-level: caller attribution is a per-application convention rather than a kernel-enforced primitive (no \texttt{position\ 4} equivalent). History-derived state is user-level: canonical per-actor state is the current actormap snapshot, with the generational chain used for time-travel and recovery rather than as the state definition. A further structural divergence: CapTP/promise pipelining vs.~external-world delegation.

\textbf{Akka Persistence} (Lightbend\cite{ref22}). Shared: state recovered by replaying an append-only event journal --- the closest ``state is history'' implementation for actors. Divergences: opt-in library on mutable actors vs.~append-only-only kernel state; application-defined events recording actor-internal mutations vs.~kernel-structured \([\mathrm{caller}, \mathrm{msg}]\) entries recording boundary crossings; no cross-actor transactions vs.~multi-object atomic bundles; infrastructure-readable journals vs.~structurally opaque logs.

\subsection{Persistence and History}

\textbf{Nock/Urbit} (Yarvin\cite{ref3,ref4}). Shared with the §5 instantiation: s-expressions over \(\mathbb{N}\), minimal instruction set. Shared at the kernel level: deterministic computation, code-is-data. Divergences: Nock is purely functional (Urbit adds persistence/identity at a higher layer; the kernel incorporates ephemeral/persistent at the kernel level); no kernel-level multi-object transaction bundles (Urbit processes events sequentially without atomic cross-process coupling); Urbit's Azimuth is external infrastructure vs.~kernel-internal opaque identity.

\textbf{Event sourcing and Datomic} (Hickey 2012\cite{ref18}, Helland 2015\cite{ref19}). Shared: state derived from an append-only event log. Datomic embodies ``state is history'' via immutable datoms with transactions as atomic bundles. Divergences: entity-attribute-value vs.~object-oriented (no object-level opacity, no per-entry provenance, no programs-with-context).

\textbf{Replicated storage with per-replica logs} (Terry et al.~1995\cite{ref26}). Shared: append-only logs, replay-to-state. Divergences: per-replica database-wide logs vs.~per-object logs; application-supplied conflict resolvers and commit-sequence-number ordering vs.~strict-serialisable kernel-transaction ordering; application-readable logs vs.~kernel-internal opacity. The kernel composes naturally with synchronous-CP wrappers (Raft, Paxos, Calvin); composing with eventually-consistent / AP-style wrappers (Bayou\cite{ref26}) loses strict serializability.

\section{Conclusion}

Our kernel as presented is a demonstration of architectural possibility, not yet a practical system. Three directions which remain open:

\begin{itemize}
\tightlist
\item
  \textbf{Surface languages}: a usable programmer-facing language that compiles to the kernel --- Self-style prototypes and Smalltalk-style classes are encodable (§7) but require concrete syntax.
\item
  \textbf{Cryptographic composition} beyond the static-allocation case sketched in §7 Case Study 2 --- Viaduct-style label tracking through dynamic object allocation is open.
\item
  \textbf{Production engineering}: §6's efficiency discussion is analytical; performance characterisation, source-to-kernel compilation, and debugging tooling remain future work.
\end{itemize}

Resource metering, distribution, consensus, and garbage collection of derived state (materialised views, snapshots, caches) are orthogonal concerns that compose with conflict-serializable scheduling (§6.3). The kernel is one foundation; commitments are explicit, the six properties hold together, and richer abstractions can be built on top.

\section{Disclosure}

We used LLM tools for literature search and copyediting assistance; all data and changes were reviewed manually.

\end{document}